\documentclass{LMCS}

\def\dOi{10(3:12)2014}
\lmcsheading%
{\dOi}
{1--18}
{}
{}
{Oct.~\phantom02, 2012}
{Aug.~27, 2014}
{}

\ACMCCS{[{\bf Theory of computation}]: Logic; Models of
  computation---Computability---Turing machines; Formal languages and
  automata theory---Automata over infinite objects}


\usepackage[latin1]{inputenc}
\usepackage{amssymb}
\usepackage{hyperref}

\def\ufootnote#1{\let\savedthfn\thefootnote\let\thefootnote\relax
\footnote{#1}\let\thefootnote\savedthfn\addtocounter{footnote}{-1}}

\newcommand{\hs}{\hspace{12mm}

}
\newcommand{\noi}{\noindent}

\newcommand{\om}{\omega}
\newcommand{\Si}{\Sigma}
\newcommand{\Sis}{\Sigma^\star}
\newcommand{\Sio}{\Sigma^\omega}
\newcommand{\nl}{\newline}

\newcommand{\fa}{\forall}
\newcommand{\ra}{\rightarrow}

\newcommand{\Ga}{\Gamma}

\newcommand{\Gao}{\Gamma^\omega}
\newcommand{\ite}{\item}

\newcommand{\ol}{$\omega$-language}

\begin{document}

\title{Ambiguity of $\om$\,-Languages of  Turing Machines} 

\author{Olivier Finkel}

\address{Equipe de Logique Math\'ematique \\Institut de Math\'ematiques de Jussieu - Paris Rive Gauche
 \\  CNRS et Universit\'e Paris 7, France.}
\email{finkel@math.univ-paris-diderot.fr}

\date{}


\begin{abstract} 
\noi An $\om$-language is a set of infinite words over a finite alphabet $X$. 
We consider the class of recursive $\om$-languages, i.e.  the class of $\om$-languages accepted by Turing machines with a B\"uchi acceptance condition, which is  also 
the class $\Si_1^1$ of (effective) analytic subsets of $X^\om$ for some finite alphabet $X$. 
We investigate here the notion of ambiguity for recursive 
$\om$-languages with regard to acceptance by B\"uchi Turing machines.  
We first present in detail essentials on the literature on
$\omega$-languages accepted by Turing Machines.   Then we give a complete and broad view on the notion of ambiguity and unambiguity of B\"uchi Turing machines and of the $\om$-languages they accept. 
To obtain our new results,  we make use of results and methods of  effective descriptive set theory. 
 \end{abstract}
\maketitle

\section{Introduction}

Languages of infinite words, also called  $\om$-languages,  accepted by finite automata were first studied by B\"uchi 
to prove the decidability of the monadic second order theory of one successor
over the integers.  Since then regular $\om$-languages have been much studied and  many applications have been  found  for specification and verification 
of non-terminating systems, 
see \cite{Thomas90,Staiger97,PerrinPin} for many results and references. 
Other  finite machines, like 
pushdown automata, multicounter automata, Petri nets, 
 have also been considered  for  reading of infinite words, 
see  \cite{Staiger97, eh, Fin-mscs06}. 

Turing  invented in  1937  what we  now call Turing machines.  
This way  he  made a unique impact on the history of computing, computer science, 
 and the mathematical theory of computability. 
Recall that the year 2012  was the Centenary of Alan Turing's birth and that many scientific events have commemorated this year  Turing's life and work.  

 The 
acceptance of infinite words by Turing machines via several acceptance conditions, like the B\"uchi or Muller ones, was studied by 
Staiger and Wagner  in   \cite{StaigerWagner77,StaigerWagner78}   and by       Cohen and Gold  in \cite{CG78b}. 
 It turned  out that 
the classes of $\om$-languages accepted by non-deterministic Turing machines with B\"uchi or Muller acceptance conditions were the same class, 
the class of effective analytic sets  \cite{StaigerWagner77,CG78b,Staiger99}.

We consider in this paper the class of recursive $\om$-languages, i.e.  the class of $\om$-languages accepted by non-deterministic Turing machines 
with a B\"uchi acceptance condition, which is  also 
the class $\Si_1^1$ of (effective) analytic subsets of $X^\om$ for some finite alphabet $X$.  

The notion of ambiguity is very important in formal language and automata theory and has been much studied for instance in the case of 
context-free finitary languages accepted by pushdown automata or generated by context-free grammars, \cite{ABB96}, and in the case of 
context-free $\om$-languages, \cite{Fin03b,Fink-Sim}.   In the case of Turing machines reading finite words,  it is easy to see that every Turing machine 
is equivalent to a deterministic, hence also unambiguous, Turing machine. 
Thus every recursive finitary language is accepted by an  unambiguous Turing machine.

We investigate here the notion of ambiguity for recursive 
$\om$-languages with regard to acceptance by B\"uchi Turing machines.  
We first present in detail essentials on the literature on
$\omega$-languages accepted by Turing Machines. In particular, we
describe the different ways of acceptance in which Turing machines
(and also other devices) might be used to accept $\omega$-languages. 
Then we give a complete and broad view on the notion of ambiguity and unambiguity of Buchi Turing machines and of the $\om$-languages they accept. 
To obtain our new results,  we make use of results and methods of  effective descriptive set theory, sometimes already used in other contexts for the study 
of other classes of $\om$-languages.

Notice that this study may first seem to be of  no practical interest, but  in fact non-deterministic
Turing machines  over infinite data seem to be relevant to real-life algorithmics over streams, where
non-determinism may appear either by choice or because of physical constraints and
perturbation.

We first show that the class of unambiguous  recursive $\om$-languages is the 
class $\Delta_1^1$ of hyperarithmetical sets.  On the other hand,   Arnold studied B\"uchi  transition systems  in  \cite{Arnold83}.  In particular,  
he proved that  the analytic subsets of 
$X^\om$ are the subsets of  $X^\om$ which are accepted by  finitely branching B\"uchi  transition systems, and that 
the Borel subsets of 
$X^\om$ are the subsets of  $X^\om$ which are accepted by  unambiguous finitely branching B\"uchi  transition systems.
Some effective versions of B\"uchi  transition systems  were studied by Staiger in 
\cite{Staiger93}. In particular, he   proved  that the subsets of  $X^\om$ which are accepted by  strictly recursive
finitely branching B\"uchi  transition systems are the effective analytic subsets of  $X^\om$.
We obtain also here  that the  $\Delta_1^1$-subsets of 
$X^\om$ are the subsets of  $X^\om$ which are accepted by   strictly   recursive unambiguous finitely branching B\"uchi  
transition systems.  This provides an effective 
analogue  to the above cited result  of  Arnold. 

Next,  we prove that recursive $\om$-languages 
satisfy  the following   dichotomy property. 
A recursive $\om$-language $L\subseteq X^\om$ is either unambiguous or has a great degree of ambiguity: for every B\"uchi 
Turing machine $\mathcal{T}$ accepting $L$, there  exist infinitely many  
$\om$-words which have $2^{\aleph_0}$ accepting runs by $\mathcal{T}$. 

We also show that if $L \subseteq  X^\om$ is accepted by a 
B\"uchi Turing machine $\mathcal{T}$ and  $L$ 
is an analytic but non-Borel set, then the set of $\om$-words, 
which have $2^{\aleph_0}$ accepting runs by $\mathcal{T}$, has cardinality $2^{\aleph_0}$. This extends a similar result of  \cite{Fink-Sim} in the 
case of context-free $\om$-languages and infinitary rational relations.
In that case we say that the  recursive $\om$-language 
$L$ has the maximum degree of ambiguity. 

Castro and Cucker studied decision problems for $\om$-languages of Turing machines in \cite{cc}. They gave the (high)  degrees
 of many classical decision problems like the emptiness, the finiteness, the cofiniteness, the universality, the equality, and  the inclusion problems. 
In \cite{Fin-HI} we  obtained many new   undecidability results about   context-free $\om$-languages and infinitary rational relations. 
We prove   here  new  undecidability results about ambiguity of recursive   $\om$-languages: 
 it is $\Pi_2^1$-complete  to determine whether a given recursive $\om$-language is 
 unambiguous and  it is $\Si_2^1$-complete to determine whether a given recursive $\om$-language has the 
 maximum degree of ambiguity.  

Then, using some recent results from \cite{Fin-ICST} and some results of set theory, 
we prove that it is equiconsistent with the axiomatic system ZFC that 
 there exists a recursive $\om$-language in the Borel class ${\bf \Pi}^0_2$, hence of low Borel rank, which has also the  
 maximum degree of ambiguity.

The paper is organized as follows. We recall some known notions in Section 2. We study unambiguous recursive $\om$-languages  in Section 3 and 
inherently  ambiguous recursive $\om$-languages  in Section 4. Some concluding remarks are given in Section 5.

\section{Reminder  of some well-known notions}
 
 We assume   the reader to be familiar with the theory of formal ($\om$-)languages  
\cite{Staiger97,PerrinPin}.
We recall the  usual notations of formal language theory. 

If  $\Si$ is a finite alphabet, a {\it non-empty finite word} over $\Si$ is any 
sequence $x=a_1\ldots a_k$, where $a_i\in\Sigma$ 
for $i=1,\ldots ,k$ , and  $k$ is an integer $\geq 1$. The {\it length}
 of $x$ is $k$, denoted by $|x|$.
 The {\it empty word} has no letter and is denoted by $\varepsilon$; its length is $0$. 
 $\Sis$  is the {\it set of finite words} (including the empty word) over $\Sigma$.
A  (finitary) {\it language} $V$ over an alphabet $\Sigma$ is a subset of  $\Sis$.

 The {\it first infinite ordinal} is $\om$.
 An $\om$-{\it word} (or infinite word) over $\Si$ is an $\om$-sequence $a_1 \ldots a_n \ldots$, where for all 
integers $ i\geq 1$, ~
$a_i \in\Sigma$.  When $\sigma=a_1 \ldots a_n \ldots$ is an $\om$-word over $\Si$, we write
 $\sigma(n)=a_n$,   $\sigma[n]=\sigma(1)\sigma(2)\ldots \sigma(n)$  for all $n\geq 1$ and $\sigma[0]=\varepsilon$.

 The  concatenation  of two finite words $u$ and $v$ is 
denoted $u\cdot v$ (and sometimes just $uv$). This operation  is extended to the product of a 
finite word $u$ and an $\om$-word $v$: the infinite word $u\cdot v$ is then the $\om$-word such that:

 $(u\cdot v)(k)=u(k)$  if $k\leq |u|$ , and 
 $(u\cdot v)(k)=v(k-|u|)$  if $k>|u|$.
  
 The {\it set of } $\om$-{\it words} over an  alphabet $\Si$ is denoted by $\Si^\om$.
An  $\om$-{\it language} $V$ over an alphabet $\Sigma$ is a subset of  $\Si^\om$, and its  complement (in $\Sio$) 
 is $\Sio - V$, denoted $V^-$.

\hs  We assume the reader to be familiar with basic notions of topology,  which
may be found in \cite{Kechris94,LescowThomas,Staiger97,PerrinPin}.
There is a natural metric on the set $\Sio$ of  infinite words 
over a finite alphabet 
$\Si$ containing at least two letters,  which is called the {\it prefix metric},  and is defined as follows. For $u, v \in \Sio$ and 
$u\neq v$ let $\delta(u, v)=2^{-l_{\mathrm{pref}(u,v)}}$ where $l_{\mathrm{pref}(u,v)}$ 
 is the first integer $n$
such that the $(n+1)^{st}$ letter of $u$ is different from the $(n+1)^{st}$ letter of $v$. 
This metric induces on $\Sio$ the usual  Cantor topology in which the {\it open subsets} of 
$\Sio$ are of the form $W\cdot \Si^\om$, for $W\subseteq \Sis$.
A set $L\subseteq \Si^\om$ is a {\it closed set} iff its complement $\Si^\om - L$ 
is an open set.

We   now recall 
the definition of the {\it Borel Hierarchy} of subsets of $X^\om$. 

\begin{defi}
For a non-null countable ordinal $\alpha$, the classes ${\bf \Si}^0_\alpha$
 and ${\bf \Pi}^0_\alpha$ of the Borel Hierarchy on the topological space $X^\om$ 
are defined as follows:
 ${\bf \Si}^0_1$ is the class of open subsets of $X^\om$, 
 ${\bf \Pi}^0_1$ is the class of closed subsets of $X^\om$, 
 and for any countable ordinal $\alpha \geq 2$: 
\nl ${\bf \Si}^0_\alpha$ is the class of countable unions of subsets of $X^\om$ in 
$\bigcup_{\gamma <\alpha}{\bf \Pi}^0_\gamma$.
 \nl ${\bf \Pi}^0_\alpha$ is the class of countable intersections of subsets of $X^\om$ in 
$\bigcup_{\gamma <\alpha}{\bf \Si}^0_\gamma$.

A set $L\subseteq X^\om$ is Borel iff it is in the union $\bigcup_{\alpha < \om_1} {\bf \Si}^0_\alpha = \bigcup_{\alpha < \om_1} {\bf \Pi}^0_\alpha$, where  
$\om_1$ is the first uncountable ordinal. 

 For 
a countable ordinal $\alpha$,  a set $L\subseteq X^\om$ is a Borel set of {\it rank} $\alpha$ iff 
it is in ${\bf \Si}^0_{\alpha}\cup {\bf \Pi}^0_{\alpha}$ but not in 
$\bigcup_{\gamma <\alpha}({\bf \Si}^0_\gamma \cup {\bf \Pi}^0_\gamma)$.

\end{defi}

\noi    
There are also some subsets of $X^\om$ which are not Borel. 
In particular, 
the class of Borel subsets of $X^\om$ is strictly included in 
the class  ${\bf \Si}^1_1$ of {\it analytic sets} which are 
obtained by projection of Borel sets. The  {\it co-analytic sets}  are the complements of 
analytic sets.  

\hs For two alphabets $X$ and $Y$ and two infinite words $x\in X^\om$ and $y\in Y^\om$, we denote 
$(x, y)$  the infinite word over the alphabet $X \times Y$ such that
$(x, y)(i)=(x(i),y(i))$ for each  integer $i\geq 1$.

\begin{defi} 
A subset $A$ of  $X^\om$ is in the class ${\bf \Si}^1_1$ of {\it analytic} sets
iff there exist a finite alphabet $Y$ and a Borel subset $B$  of  $(X \times Y)^\om$ 
such that:  $[ x \in A ] \Longleftrightarrow [ \exists y \in Y^\om $  $(x, y) \in B ]$. 
\end{defi} 

We now define completeness with regard to reduction by continuous functions. 
For a countable ordinal  $\alpha\geq 1$, a set $F\subseteq X^\om$ is said to be 
a ${\bf \Si}^0_\alpha$  
(respectively,  ${\bf \Pi}^0_\alpha$, ${\bf \Si}^1_1$)-{\it complete set} 
iff for any set $E\subseteq Y^\om$  (with $Y$ a finite alphabet): 
 $E\in {\bf \Si}^0_\alpha$ (respectively,  $E\in {\bf \Pi}^0_\alpha$,  $E\in {\bf \Si}^1_1$) 
iff there exists a continuous function $f: Y^\om \ra X^\om$ such that $E = f^{-1}(F)$, i.e. such that ($x\in E$ iff $f(x)\in F$).

\hs We now  recall  the definition of the  arithmetical hierarchy of  \ol s which form the effective analogue to the 
hierarchy of Borel sets of finite ranks, see \cite{Staiger97}.

 Let $X$ be a finite alphabet. An \ol~ $L\subseteq X^\om$  belongs to the class 
$\Si_n$ if and only if there exists a recursive relation 
$R_L\subseteq (\mathbb{N})^{n-1}\times X^\star$  such that
$$L = \{\sigma \in X^\om \mid \exists k_1\ldots Q_nk_n  \quad (k_1,\ldots , k_{n-1}, 
\sigma[k_n+1])\in R_L \}$$

\noi where $Q_i$ is one of the quantifiers $\fa$ or $\exists$ 
(not necessarily in an alternating order). An \ol~ $L\subseteq X^\om$  belongs to the class 
$\Pi_n$ if and only if its complement $X^\om - L$  belongs to the class 
$\Si_n$.  The inclusion relations that hold  between the classes $\Si_n$ and $\Pi_n$ are 
the same as for the corresponding classes of the Borel hierarchy. 
 The classes $\Si_n$ and $\Pi_n$ are  included in the respective classes 
${\bf \Si}_n^0$ and ${\bf \Pi}_n^0$ of the Borel hierarchy, and cardinality arguments suffice to show that these inclusions are strict. 

  As in the case of the Borel hierarchy, projections of arithmetical sets 
 lead 
beyond the arithmetical hierarchy, to the analytical hierarchy of \ol s. The first class 
of this hierarchy is the (lightface) class $\Si^1_1$ of {\it effective analytic sets} 
 which are obtained by projection of arithmetical sets.
 In fact an \ol~ $L\subseteq X^\om$  is in the class $\Si_1^1$ iff it is the projection 
of an \ol~ over the alphabet $X\times \{0, 1\}$ which is in the class $\Pi_2$.  The (lightface)  class $\Pi_1^1$ of  {\it effective co-analytic sets} 
 is simply the class of complements of effective analytic sets. We denote as usual $\Delta_1^1 = \Si^1_1 \cap \Pi_1^1$.  

The (lightface) class $\Si_1^1$ of effective analytic sets is strictly included into the (boldface) class ${\bf \Si}^1_1$ of analytic sets. 

We assume the reader to be familiar with the  arithmetical and  analytical hierarchies
on subsets of  $\mathbb{N}$, these notions  may be found in the textbooks on computability theory \cite{rog} 
\cite{Odifreddi1,Odifreddi2}.  Notice that we will not have to consider subsets of $\mathbb{N}$ of ranks greater than 2 in the analytical hierarchy, 
so the most complex subsets of $\mathbb{N}$  occcuring in this paper will be $\Si^1_2$-sets or $\Pi^1_2$-sets. 

We shall  consider in the sequel some $\Si_1^1$ or $\Pi_1^1$ subsets of product spaces like $X^\om \times Y^\om$ or $\mathbb{N}\times Y^\om$. 
Moreover,  in effective descriptive set theory one often  considers the notion of relativized class  $\Si_1^1(w)$:  for  $w\in X^\om$, 
a set $L \subseteq Y^\om$ is a $\Si_1^1(w)$-set iff there exists a $\Si_1^1$-set $T \subseteq  X^\om \times Y^\om$ 
such that $L=\{ y\in Y^\om \mid (w, y) \in T \}$. A set $L \subseteq Y^\om$ is a $\Pi_1^1(w)$-set iff its complement is a $\Si_1^1(w)$-set. 
A set $L \subseteq Y^\om$ is a $\Delta_1^1(w)$-set iff it is in the class $\Si_1^1(w) \cap \Pi_1^1(w)$. 
We say that  $y \in Y^\om$ is in the class 
$\Delta_1^1$ (respectively, $\Delta_1^1(w)$) iff the singleton $\{y\}$ is a $\Delta_1^1$-set (respectively, $\Delta_1^1(w)$-set). 

\hs Recall now  the notion of acceptance of infinite words by Turing machines considered  by  Cohen and Gold  in \cite{CG78b}. 

\begin{defi}
A non-deterministic Turing machine $\mathcal{M}$ is a $5$-tuple $\mathcal{M}=(Q, \Si, \Ga, \delta, q_0)$, where $Q$ is a finite set of states, 
$\Si$ is a finite input alphabet, $\Ga$ is a finite tape alphabet satisfying $\Si  \subseteq \Ga$, $q_0$ is the initial state, 
and $\delta$ is a mapping from $Q \times \Ga$ to subsets of $Q \times \Ga \times \{L, R, S\}$. A configuration of $\mathcal{M}$ is a triplet 
$(q, \sigma, i)$, where $q\in Q$, $\sigma \in \Ga^\om$ and $i\in \mathbb{N}$. An infinite sequence of configurations $r=(q_i, \alpha_i, j_i)_{i\geq 1}$
is called a run of $\mathcal{M}$ on $w\in \Sio$ iff: 
\begin{enumerate}[label=\({\alph*}]
\ite $(q_1, \alpha_1, j_1)=(q_0, w, 1)$, and 
\ite for each $i\geq 1$, $(q_i, \alpha_i, j_i) \vdash (q_{i+1}, \alpha_{i+1}, j_{i+1})$, 
\end{enumerate}
\noi where $\vdash$ is the transition relation of $\mathcal{M}$ defined as usual. The run $r$ is said to be complete if every position is visited, i.e. if 
$(\fa n \geq 1) (\exists k \geq 1) (j_k \geq n)$. The run $r$ is said to be oscillating if some position is visited infinitely often, i.e. if 
$(\exists k \geq 1) (\fa n \geq 1) (\exists m \geq n) ( j_m=k)$. 
\end{defi}

\begin{defi}
Let $\mathcal{M}=(Q, \Si, \Ga, \delta, q_0)$ be a non-deterministic Turing machine   
and $F \subseteq Q$, $\mathcal{F}\subseteq 2^Q$. The $\om$-language $1'$-accepted (respectively, $2$-accepted)   by $(\mathcal{M}, F)$ is 
the set of $\om$-words $ \sigma \in \Sio$ such that there exists a complete non-oscillating run $ r=(q_i, \alpha_i, j_i)_{i\geq 1}$
 of $\mathcal{M}$  on  $\sigma$ such that, for all $ i, q_i \in F$  (respectively, for infinitely many $ i, q_i \in F$). The $\om$-language $3$-accepted 
by $(\mathcal{M}, \mathcal{F})$ is the  set of $\om$-words $ \sigma \in \Sio$ such that there exists a complete non-oscillating run $ r$
 of $\mathcal{M}$  on  $\sigma$ such that the set of states appearing infinitely often during the run $r$ is an element of $\mathcal{F}$. 
\end{defi}

\noi The $1'$-acceptance condition  is also considered  by Castro and Cucker in \cite{cc}. 
The $2$-acceptance and $3$-acceptance  conditions are now usually called B\"uchi and  Muller
acceptance conditions. 
 Cohen and Gold proved the following result  in \cite[Theorem 8.2]{CG78b}. 

\begin{thm}[Cohen and Gold  \cite{CG78b}]
An $\om$-language is accepted by a non-deterministic Turing machine with 
$1'$-acceptance condition iff it is accepted by a non-deterministic Turing machine with B\"uchi (or  Muller) acceptance condition. 
\end{thm}

\noi 
Notice that this result holds 
because  Cohen and Gold's Turing  machines accept infinite words via {\it complete non-oscillating runs}, while  $1'$, B\"uchi or Muller acceptance conditions
 refer to the sequence of states entered during an infinite run. 

\hs There are actually three types of a required behaviour on the input tape which have been considered in the literature. We now recall the classification 
of these  three types given in \cite{Staiger99,Staiger00}. 

\hs {\bf  Type 1.}  This is the type considered in  \cite{StaigerWagner77,StaigerWagner78,Staiger97}.  Here we do not take into consideration 
the behaviour of the Turing machine on the input tape. Thus the acceptance depends only on the infinite sequence of states entered by the machine  during the 
infinite computation. In particular, the machine may not read the whole input tape. 

\hs {\bf  Type 2.}  This is the appoach of  \cite{eh}. Here one requires  that the machine reads the whole infinite tape (i.e. that the run is complete).  

\hs {\bf  Type 3.}  This is the type which is considered by Cohen and Gold in \cite{CG78b};  the acceptance of infinite words is 
defined   via {\it complete non-oscillating runs}. 

\hs We refer to   \cite{StaigerWagner78, Staiger99, StaigerFreund99, Staiger00} 
for a study  of these different approaches. They are in particular explicitely investigated for deterministic Turing machines in  \cite{StaigerWagner78}, 
\cite{Staiger99}, and \cite{StaigerFreund99}. 

\hs Notice that ``reading the whole input tape" solely  is  covered  by the B\"uchi acceptance condition. 

In this paper,  we shall consider Turing machines accepting  $\om$-words via acceptance by  runs reading the whole input tape 
(i.e., not necessarily non-oscillating). 
 By  \cite[Theorem 16]{Staiger99} (see also \cite[Theorem 5.2]{Staiger00})  
we have the following characterization of the class of   $\om$-languages accepted by these non-deterministic 
Turing machines. 

\begin{thm}[\cite{Staiger99}]\label{s}
The  class of  $\om$-languages 
accepted by non-deterministic Turing machines with  $1'$ (respectively, B\"uchi,   Muller)  acceptance condition 
is  the class $\Sigma_1^1$ of  effective analytic sets.
\end{thm}

In the sequel we shall also restrict our study to the B\"uchi acceptance condition. But one can  easily see that all the results of this paper are  true for 
any other acceptance condition leading to the class $\Sigma_1^1$ of  effective analytic sets. For instance 
it  follows from \cite[Note 2 page 12]{CG78b}  and from   Theorem \ref{s} that the  class of  $\om$-languages 
accepted by Cohen's and Gold's non-deterministic Turing machines with  $1'$ (respectively, B\"uchi,   Muller) acceptance condition 
is  the class $\Sigma_1^1$. Moreover the class $\Sigma_1^1$ is also the class of $\om$-languages accepted by Turing machines with B\"uchi 
acceptance condition if we do not  require that  the Turing machine reads the whole infinite tape but only that it 
runs forever,  \cite{Staiger97}. 

Due to the above results, we shall say, as in \cite{Staiger97},  that  an $\om$-language is {\it recursive} iff it belongs to 
 the class $\Sigma_1^1$. Notice that in another presentation, 
as in \cite{rog}, the recursive $\om$-languages are those which are in the class $\Si_1 \cap \Pi_1$, see also \cite{LescowThomas}. 

On the other hand, we mention that $\om$-languages of deterministic Turing machines form the class of boolean combinations 
of arithmetical $\Pi_2^0$-sets, \cite{Staiger97}.
Selivanov gave a very fine topological classification of these languages, based on the Wadge hierarchy of Borel sets,   in 
\cite{Selivanov03a,Selivanov03b}.

\section{Unambiguous recursive  $\om$-languages}

We have said  in the preceding section that we shall  restrict our study to the B\"uchi acceptance 
condition and to acceptance via runs reading the whole input tape.  

\hs We now briefly justify the restriction to Type 2 acceptance  for the study of ambiguity of recursive $\om$-languages, by showing 
 that  the three types defined in the preceding section, along with the B\"uchi acceptance 
condition, give the same class of $\om$-languages accepted by unambiguous  Turing machines. 

\hs We first give the two definitions. 

\begin{defi}\label{31}
A B\"uchi Turing machine $\mathcal{M}$ with Type $i$ acceptance,  reading $\om$-words over an alphabet $\Si$,  is said to 
be unambiguous iff for every $\om$-word $x\in \Sio$  the machine $\mathcal{M}$ has at most one 
accepting run over $x$. 
\end{defi}

\begin{defi}\label{32}
Let $\Si$ be a finite alphabet. A recursive $\om$-language $L \subseteq  \Sio$  is said to be unambiguous of Type $i$ 
iff  it is accepted by (at least) one unambiguous 
B\"uchi Turing machine with Type $i$ acceptance. Otherwise the recursive $\om$-language $L$ is said to be inherently ambiguous of Type $i$. 
\end{defi}

We now informally explain why an $\om$-language is unambiguous for Type 1 acceptance iff it is unambiguous for Type 2 acceptance iff it is 
unambiguous for Type 3 acceptance.

\hs \noi {\bf (Type 1 unambiguity)$\Rightarrow$ (Type 2 unambiguity)}.

Let $L$ be an $\om$-language which is accepted by an unambiguous B\"uchi Turing machine $\mathcal{M}$ for Type 1 acceptance. Using the ``Folding process" described by Cohen and Gold in 
\cite[pages 11-12]{CG78b}, we can construct another Turing machine $\mathcal{M}'$ which simulates the machine $\mathcal{M}$ and accepts the same language but 
which has only complete and non-oscillating runs. Notice that each infinite run of $\mathcal{M}$ provides a unique  run of $\mathcal{M}'$ thus  the $\om$-language $L$ is accepted 
unambiguously by the B\"uchi Turing machine $\mathcal{M}'$ for Type 2 (and also Type 3) acceptance. 

\hs \noi {\bf (Type 2 unambiguity)$\Rightarrow$ (Type 3 unambiguity)}.

Let $L$ be an $\om$-language which is accepted by an unambiguous B\"uchi Turing machine $\mathcal{M}$ for Type 2 acceptance. Using the fact that 
``reading the whole input tape" solely  is  covered  by the B\"uchi acceptance condition, one can construct an equivalent Type 2 B\"uchi  Turing machine $\mathcal{M}'$ which is still unambiguous 
and accepts the same language with the following additional property: any run which is not complete does not satisfy the B\"uchi condition. Next we can  use again the ``Folding process" (see 
\cite[Note 2 page 12]{CG78b}) and obtain an unambiguous B\"uchi Turing machine $\mathcal{M}''$ for Type 3 acceptance which accepts the same   $\om$-language $L$.

\hs \noi {\bf (Type 3 unambiguity)$\Rightarrow$ (Type 1 unambiguity)}. 

Let $L$ be an $\om$-language which is accepted by an unambiguous B\"uchi Turing machine $\mathcal{M}$ for Type 3 acceptance. Then every $\om$-word $x$ which is accepted by the machine 
$\mathcal{M}$ has a unique accepting run. But there may exist some non-complete, or oscillating, runs of $\mathcal{M}$ over $x$ which satisfy the  B\"uchi  acceptance  condition. 
Intuitively we can transform the machine $\mathcal{M}$ to obtain a new machine $\mathcal{M}'$ which has essentially the same runs but in such a way that non-complete, or oscillating, 
runs of $\mathcal{M}'$ will no longer satisfy the  B\"uchi  acceptance  condition. Then the new  B\"uchi Turing machine $\mathcal{M}'$  accepts the same $\om$-language 
$L$ but for Type 1 acceptance and the machine $\mathcal{M}'$ is unambiguous.

\hs From now on in this paper  a B\"uchi Turing machine will be  a Turing machine reading $\om$-words and accepting $\om$-words with a 
B\"uchi acceptance condition via  runs reading the whole input tape. And we shall say that a recursive $\om$-language is unambiguous iff it is unambiguous 
of Type $2$ (iff it is unambiguous of Type 1 or 3).

\hs  We can now state our first result. 

\begin{prop}\label{p1}
If  $\Si$ is a finite alphabet and  $L \subseteq  \Sio$  is an  unambiguous recursive $\om$-language then $L$ belongs to the (effective) class $\Delta_1^1$. 
\end{prop}

\proof Let $L \subseteq  \Sio$ be an $\om$-language accepted by an unambiguous 
B\"uchi Turing machine $(\mathcal{M}, F)$, where $\mathcal{M}=(Q, \Si, \Ga, \delta, q_0)$ is  a Turing machine   
and $F \subseteq Q$. Recall that a configuration of the Turing machine $\mathcal{M}$ is a triple $(q, \sigma, i)$, 
where $q\in Q$, $\sigma \in \Ga^\om$ and $i\in \mathbb{N}$. It can be coded by the infinite word $q^{i}\cdot \sigma$ over the alphabet 
$Q \cup \Ga$,  where we have assumed without loss of generality that $Q$ and $\Ga$ are disjoint. Then a run  of  $\mathcal{M}$ on $w\in \Sio$ is 
an infinite sequence of configurations $r=(q_i, \alpha_i, j_i)_{i\geq 1}$ which is then coded by an infinite sequence of $\om$-words 
$(r_i)_{i\geq 1}=(q_i^{j_i}\cdot \alpha_i)_{i\geq 1}$ over $Q \cup \Ga$. 
Using now a recursive bijection $b: (\mathbb{N}\setminus \{ 0 \})^2 \ra \mathbb{N}\setminus \{ 0 \}$
 and its inverse $b^{-1}$  we can  effectively code the sequence 
$(r_i)_{i\geq 1}$ by a single infinite word  $r'\in (Q \cup \Ga)^\om$ defined by: for every  integer $j\geq 1$ such that $b^{-1}(j)=(i_1, i_2)$, 
 $r'(j)=r_{i_1}(i_2)$. Moreover the infinite word  $r'\in (Q \cup \Ga)^\om$ can be coded in a recursive manner 
by an infinite word over the alphabet $\{0, 1\}$. 
We can then identify $r$ with its code $\bar{r}\in \{0, 1\}^\om$ and this will be often done in the sequel. 
  Let now $R$ be defined  by: 
$$R= \{ (w, r) \mid w\in \Si^\om \mbox{ and }  r \in \{0, 1\}^\om  \mbox{ is an accepting run of } (\mathcal{M}, F)  \mbox{ on  the $\om$-word } w \}.$$
\noi The set $R$ is  a $\Delta^1_1$-set, and even    an arithmetical set:  it is easy to see that it is accepted by a {\it deterministic} Muller  Turing machine
 and thus  it is a $\Delta_3^0$-subset of the space  $(\Si\times \{0, 1\})^\om$, see \cite{Staiger97}.  

\hs Consider now the projection $\mathrm{PROJ}_{\Si^\om} : ~ \Si^\om \times \{0, 1\}^\om \ra \Si^\om$ defined by 
$\mathrm{PROJ}_{\Si^\om}(w, r) = w$ for all $(w, r) \in  \Si^\om \times  \{0, 1\}^\om$. This projection is a recursive  function, i.e. 
``there is an algorithm which given sufficiently close approximations to $(w, r)$ produces arbitrarily accurate approximations to
 $\mathrm{PROJ}_{\Si^\om}(w, r)$", see  \cite{Moschovakis09}.   
Moreover  it is {\it injective} on the  $\Delta^1_1$-set $R$ because the B\"uchi Turing machine $(\mathcal{M}, F)$ is unambiguous. 
But the image of a $\Delta^1_1$-set by an injective recursive  function is  a $\Delta^1_1$-set, see \cite[page 169]{Moschovakis09}
and thus the  recursive $\om$-language  $L=\mathrm{PROJ}_{\Si^\om}(R)$ is a $\Delta^1_1$-subset of $\Si^\om$. 
\qed 

\hs In order to prove a converse statement we now  first recall the notion of  B\"uchi transition system. 

\begin{defi}
A  B\"uchi transition system is a tuple $\mathcal{T}=(\Si, Q, \delta, q_0, Q_f)$, where $\Si$ is a finite input alphabet, $Q$ is a countable set of states, 
$\delta \subseteq Q \times \Si  \times Q$ 
is the transition relation, $q_0 \in Q$ is the initial state, 
and $Q_f \subseteq Q$ is the set of final states. 
A run of $\mathcal{T}$ over an infinite word $\sigma \in \Sio$ is an infinite sequence 
of states $(t_i)_{i\geq 0}$, such that 
~ $t_0 = q_0$, ~and  for each $i\geq 0$,  ~ $(t_i, \sigma (i+1), t_{i+1}) \in \delta$. The run is said to be accepting iff there are infinitely 
many integers $i$ such that $t_i$ is in $Q_f$. An $\om$-word $\sigma \in \Sio$ is accepted by  $\mathcal{T}$ iff there is (at least) one accepting run
of $\mathcal{T}$ over $\sigma$. The $\om$-language $L(\mathcal{T})$ accepted by $\mathcal{T}$ is the set of  $\om$-words accepted by  $\mathcal{T}$. 
The transition system is said to be $unambiguous$ if each infinite word $\sigma \in \Sio$ has at most 
one accepting run by $\mathcal{T}$.
The transition system is said to be $finitely\ branching$ if for each state $q\in Q$ and each $a \in \Si$, there are only finitely 
many states $q'$ such that $(q, a, q') \in \delta$. 
\end{defi}

Arnold proved the following theorem in \cite{Arnold83}. 

\begin{thm}\label{ts}
Let $\Si$ be an alphabet having at least two letters. 
\begin{enumerate}
\ite 
 The analytic subsets of 
$\Sio$ are the subsets of  $\Sio$ which are accepted by  finitely branching B\"uchi  transition systems.
\ite 
The Borel subsets of 
$\Sio$ are the subsets of  $\Sio$ which are accepted by  unambiguous finitely branching B\"uchi  transition systems.
\end{enumerate}
\end{thm}

\noi It is also very natural to consider effective versions of B\"uchi  transition systems where the sets $Q, \delta,$ and $Q_f$ are recursive. Such transition systems 
are studied by Staiger in \cite{Staiger93} where $Q$ is actually either the set $\mathbb{N}$ of natural numbers or a finite segment of it,  and 
they are called {\it strictly recursive}. It is proved by Staiger that the subsets of  $\Sio$ which are accepted by  strictly recursive
finitely branching B\"uchi  transition systems are the effective analytic subsets of  $\Sio$.

On the other hand, the B\"uchi  transition systems are considered by Finkel and Lecomte in \cite{Fink-Lec2} where 
 they are used in the study of topological properties of $\om$-powers.  Using an effective version of a theorem of  Kuratowski, it is proved in 
 \cite{Fink-Lec2} that every $\Delta^1_1$-subset of $\{0, 1\}^\om$ is actually accepted by an unambiguous strictly   recursive finitely branching 
B\"uchi  transition system (where the degree of branching of the transition system is actually equal to 2). 
Using an easy coding  this is easily extended to the case of any $\Delta^1_1$-subset of $\Si^\om$, where $\Si$ is a finite alphabet.

\begin{thm}\label{un}
 Let $\Si$ be an alphabet having at least two letters. 
An $\om$-language  $L \subseteq  \Sio$  is an  unambiguous recursive $\om$-language iff  $L$ belongs to the (effective) class $\Delta_1^1$. 
\end{thm}

\proof The implication from left to right is given by Proposition \ref{p1}.  We now  prove the reverse implication. 
Using the fact that every recursive set of finite words over a finite alphabet  $\Ga$  
is accepted by a {\it deterministic} hence also unambiguous Turing machine reading finite words, we can easily see that 
every $\om$-language which is accepted by an unambiguous strictly   recursive finitely branching 
B\"uchi  transition system is also accepted by an unambiguous    B\"uchi     Turing machine.
\qed

\hs Notice that we have also the effective analogue to Arnold's Theorem \ref{ts}. 

\begin{thm}\label{ts2}
Let $\Si$ be an alphabet having at least two letters. 
\begin{enumerate}
\ite 
 The effective analytic subsets of 
$\Sio$ are the subsets of  $\Sio$ which are accepted by  strictly   recursive  finitely branching B\"uchi  transition systems.
\ite 
The $\Delta_1^1$-subsets of 
$\Sio$ are the subsets of  $\Sio$ which are accepted by   strictly   recursive unambiguous finitely branching B\"uchi  transition systems.
\end{enumerate}
\end{thm}

\proof Item 1 is proved in \cite{Staiger93}.  To prove that every  $\om$-language which is   accepted by a
  strictly   recursive unambiguous finitely branching B\"uchi  transition system is a  $\Delta_1^1$-set we can reason as in the case of Turing machines 
(see the proof of Proposition \ref{p1}). As said above, the  converse statement is proved in \cite{Fink-Lec2}. 
\qed

\section{Inherently ambiguous recursive  $\om$-languages}

The notion of ambiguity for context-free $\om$-languages has been studied in \cite{Fin03b,Fink-Sim}.  In particular it was proved in 
\cite{Fink-Sim} that every  context-free $\om$-language which is non-Borel has a maximum degree of ambiguity. This was proved by stating firstly  
a lemma,  using a theorem of Lusin and Novikov. We now recall this lemma and its proof. 

\begin{lem}[\cite{Fink-Sim}]\label{lem2}
Let $\Si$ and $X$ be two finite alphabets having at least two letters and 
 $B$ be a Borel subset of 
$\Sio \times X^\om$ such that $\mathrm{PROJ}_{\Sio}(B)$ is not a Borel subset of $\Sio$.
Then there are $2^{\aleph_0}$ $\om$-words  $\alpha \in \Sio$ such that the section $B_\alpha=\{\beta \in X^\om \mid (\alpha,\beta) \in B\}$ 
has cardinality $2^{\aleph_0}$.
\end{lem}

\proof Let $\Si$ and $X$ be two finite alphabets having at least two letters and 
 $B$ be a Borel subset of 
$\Sio \times X^\om$ such that $\mathrm{PROJ}_{\Sio}(B)$ is not  Borel.  
 In a first step we  prove that 
there are uncountably many $\alpha \in \Sio$ such that the section $B_\alpha$ 
is uncountable.
 Recall that by a Theorem of Lusin and Novikov, see \cite[page 123]{Kechris94}, if for
 all  $\alpha \in \Sio$, the section $B_\alpha$ of the Borel set $B$ was   countable,  
 then  $\mathrm{PROJ}_{\Sio}(B)$ would be  a Borel subset of $\Sio$. 
 Thus there exists at least  one $\alpha \in \Sio$ such that  $B_\alpha$  is uncountable.
In fact  we can prove that  the set $U=\{\alpha \in \Sio \mid  B_\alpha  
\mbox{ is uncountable } \}$ is uncountable, otherwise 
$U=\{\alpha_0, \alpha_1, \ldots \alpha_n, \ldots \}$ would be Borel as the countable union 
of the closed sets $\{\alpha_i\}$, $i\geq 0$. 
Notice that for $\alpha \in \Sio$ we have 
$\{\alpha\} \times B_\alpha = B \cap [ \{\alpha\} \times X^\om ]$ so the set $\{\alpha\} \times B_\alpha$ is Borel 
 as intersection of two Borel sets. 
 Thus for each $n\geq 0$ the set $\{\alpha_n\}\times B_{\alpha_n}$ would be  Borel, 
and  $C=\cup_{n \in \om}\{\alpha_n\}\times B_{\alpha_n}$ would be Borel as a 
countable union of Borel sets. 
So $D=B - C$ would be borel too. 
 But all sections of $D$ would be countable thus 
 $\mathrm{PROJ}_{\Sio}(D)$ would be Borel by Lusin and Novikov's Theorem. 
 Then $\mathrm{PROJ}_{\Sio}(B)= U \cup \mathrm{PROJ}_{\Sio}(D)$ would be 
also Borel as union of two Borel sets, and this would lead to a
contradiction.  So we have proved that the set 
$\{ \alpha \in \Sio \mid B_\alpha \mbox{ is uncountable } \}$
 is uncountable. 

 On the other hand we know from 
another Theorem of Descriptive Set Theory that the set 
$\{ \alpha \in \Sio \mid B_\alpha \mbox{ is countable } \}$ is a ${\bf \Pi^1_1}$-subset of 
$\Sio$, see \cite[page 123]{Kechris94}.  
 Thus its complement $\{ \alpha \in \Sio \mid B_\alpha \mbox{ is uncountable } \}$
is analytic. 
But by Suslin's Theorem an analytic subset of   $\Sio$ is either countable 
or has cardinality $2^{\aleph_0}$, \cite[p. 88]{Kechris94}. Therefore the set 
$\{ \alpha \in \Sio \mid B_\alpha \mbox{ is uncountable } \}$
has cardinality $2^{\aleph_0}$. 
 Recall now  that we have already seen that, for each $\alpha \in \Sio$, the set 
 $\{\alpha\} \times B_\alpha$ is Borel. 
Thus  $B_\alpha$ 
itself is Borel and  by Suslin's Theorem $B_\alpha$ is either countable or has cardinality $2^{\aleph_0}$.
From this we deduce that 
$\{ \alpha \in \Sio \mid B_\alpha \mbox{ is uncountable } \} = 
\{ \alpha \in \Sio \mid B_\alpha \mbox{ has cardinality } 2^{\aleph_0} \}$ 
has cardinality $2^{\aleph_0}$. \qed

\hs We can now apply  this lemma to the study of ambiguity of Turing machines, in a similar way as in \cite{Fink-Sim} for context-free $\om$-languages. 
We can now state  the following result. 

\begin{thm}\label{mainthe}
Let $L \subseteq  \Sio$ be an $\om$-language accepted by a 
B\"uchi Turing machine $(\mathcal{M}, F)$   such that $L$ 
is an analytic but non-Borel set. The set of $\om$-words, 
which have $2^{\aleph_0}$ accepting runs by $(\mathcal{M}, F)$, has cardinality $2^{\aleph_0}$. 
\end{thm}

\proof Let $L \subseteq  \Sio$ be an analytic but non-Borel  $\om$-language accepted by a
B\"uchi Turing machine $(\mathcal{M}, F)$, where $\mathcal{M}=(Q, \Si, \Ga, \delta, q_0)$ is  a Turing machine   
and $F \subseteq Q$.  As in the proof of Proposition \ref{p1} we consider the set 
$R$  defined  by: 
$$R= \{ (w, r) \mid w\in \Si^\om \mbox{ and }  r \in \{0, 1\}^\om  \mbox{ is an accepting run of } (\mathcal{M}, F)  \mbox{ on  the $\om$-word } w \}.$$
\noi The set $R$ is  a $\Delta^1_1$-set, and thus it is a Borel subset of    $ \Si^\om \times  \{0, 1\}^\om$. But by hypothesis the set 
$\mathrm{PROJ}_{\Si^\om}(R)=L$ is not Borel. Thus it follows from Lemma \ref{lem2} that the set of $\om$-words, 
which have $2^{\aleph_0}$ accepting runs by $(\mathcal{M}, F)$, has cardinality $2^{\aleph_0}$. 
\qed 

\hs  We now know that every recursive $\om$-language which is non-Borel has a maximum degree of ambiguity. 
On the other hand Proposition \ref{p1} states that 
every recursive $\om$-language which does not belong to the (effective) class $\Delta_1^1$ is actually inherently ambiguous.  In fact we can prove a stronger 
result, using the following effective version of a theorem of Lusin and Novikov: 

\begin{thm}[see 4.F.16   page 195  of   \cite{Moschovakis09}]\label{eff-ls}
 Let $\Si$ and $X$ be two finite alphabets having at least two letters and 
 $B$ be a     $\Delta_1^1$-subset of 
$\Sio \times X^\om$ such that for all  $\alpha \in \Sio$ the section $B_\alpha=\{\beta \in X^\om \mid (\alpha,\beta) \in B\}$ is countable. Then 
the set $\mathrm{PROJ}_{\Sio}(B)$ is also a $\Delta_1^1$-subset of  $\Sio$. 
\end{thm}

We can now state the following result. 

\begin{thm}\label{inf}
Let $L \subseteq  \Sio$ be an $\om$-language accepted by a 
B\"uchi Turing machine $(\mathcal{M}, F)$   such that $L$ 
is not a $\Delta_1^1$-set. Then there exist infinitely many  $\om$-words 
which have $2^{\aleph_0}$ accepting runs by $(\mathcal{M}, F)$. 
\end{thm}

\proof Let $L \subseteq  \Sio$ be an    $\om$-language which is not a $\Delta_1^1$-set and which is accepted by a
B\"uchi Turing machine $(\mathcal{M}, F)$, where $\mathcal{M}=(Q, \Si, \Ga, \delta, q_0)$ is  a Turing machine   
and $F \subseteq Q$.  As in the proof of Proposition \ref{p1} we consider the set 
$R$  defined  by: 
$$R= \{ (w, r) \mid w\in \Si^\om \mbox{ and }  r \in \{0, 1\}^\om  \mbox{ is an accepting run of } (\mathcal{M}, F)  \mbox{ on  the $\om$-word } w \}.$$
\noi The set of accepting runs of $(\mathcal{M}, F)$ on  an $\om$-word $w\in \Si^\om$ is the section 
$$R_w=\{  r \in \{0, 1\}^\om \mid (w, r) \in R\}.$$ 
\noi  We have seen that the set $R$ is a $\Delta_1^1$-set hence also a $\Si_1^1$-set, 
and thus for each  $\om$-word $w\in \Si^\om$ the set $R_w$ is in the 
relativized class $\Sigma_1^1(w)$.  On the other hand it is known that a $\Sigma_1^1(w)$-set is countable if and only if all of its members are in the class 
$\Delta_1^1(w)$, see   \cite[page 184]{Moschovakis09}. Therefore the set $R_w$ is countable iff for all $r\in R_w$ ~ $r\in \Delta_1^1(w)$. 
Notice also that $R_w$ is an analytic set thus it is either countable or has the cardinality $2^{\aleph_0}$ of the continuum. 

\hs Recall that Harrington, Kechris and Louveau obtained  a coding of $\Delta_1^1$-subsets  of  $\{0, 1\}^\om$  in \cite{HKL}.
Notice that in the same way they obtained also a coding of the   $\Delta_1^1(w)$-subsets  of  $\{0, 1\}^\om$ which we now recall. 

 For each $w\in  \Si^\om$ there esists a $\Pi_1^1(w)$-set $W(w) \subseteq \mathbb{N}$ and a 
$\Pi_1^1(w)$-set $C(w) \subseteq \mathbb{N}\times \{0, 1\}^\om$ such that,  if we denote $C_n(w) = \{x\in \{0, 1\}^\om \mid (n, x) \in C(w) \}$, then 
 $\{(n, \alpha) \in \mathbb{N}\times \{0, 1\}^\om \mid  n\in W(w) \mbox{ and } \alpha \notin C_n(w) \}$ is a $\Pi_1^1(w)$-subset 
of the product space $ \mathbb{N}\times \{0, 1\}^\om$ and the $\Delta_1^1(w)$-subsets of $\{0, 1\}^\om$ are the sets of the form $C_n(w)$ for $n\in W(w)$. 

\hs We can  now   express $[ (\exists n \in W(w) )~  C_n(w)=\{x\} ]$  by the sentence $\phi(x,w)$:  
\[\eqalign{\exists n ~[ ~ &n \in W(w)   \mbox{ and }   (n,x) \in C(w)
    \mbox{ and }\cr& \fa y \in  \{0, 1\}^\om~ [ ( n\in W(w)  \mbox{ and
      } (n,y) \notin C(w) )  
\mbox{ or } ( y=x ) ] ]
  }
\]

\noi But we know that $C(w)$ is a $\Pi_1^1(w)$-set and that  
$\{(n, \alpha) \in \mathbb{N}\times \{0, 1\}^\om \mid  n\in W(w) \mbox{ and } \alpha \notin C_n(w) \}$ is a $\Pi_1^1(w)$-subset 
of  $ \mathbb{N}\times \{0, 1\}^\om$. Moreover the quantification  $\exists n $ in the above formula is a first-order quantification therefore the above formula 
$\phi(x,w)$ is a  $\Pi_1^1$-formula. 
We can now  express that $R_w \mbox{  is countable}$ by the  sentence $\psi(w)$ : 
$$\fa x \in \{0, 1\}^\om ~~ [ (  x \notin  R_w  ) \mbox{ or } ( \exists n \in W(w) ~  C_n(w)=\{x\} ) ]$$
that is, 
$$\fa x \in \{0, 1\}^\om ~~ [ (  x \notin  R_w ) \mbox{ or } \phi(x,w)  ]$$
\noi This is a $\Pi_1^1$-formula thus $R_w \mbox{  is uncountable}$ is expressed by a $\Si_1^1$-formula and thus the set 
$$D=\{  w \mid w\in \Si^\om \mbox{ and there are  uncountably many  accepting runs of } (\mathcal{M}, F)  \mbox{ on } w \}.$$
\noi is a  $\Si_1^1$-set. 

\hs Towards a contradiction, assume now that the set $D$ is finite. Then for every $x\in D$ the singleton $\{x\}$ is a $\Delta_1^1$-subset of  $\{0, 1\}^\om$
because $D$ is a countable $\Si_1^1$-set. But $D$ is finite so it would be the union of a {\it finite } set of $\Delta_1^1$-sets and thus it would be also a 
$\Delta_1^1$-set. Consider now the set $R'=R \setminus ( D\times \{0, 1\}^\om )$. This set would be also a $\Delta_1^1$-set and  
$\mathrm{PROJ}_{\Sio}(R')=L\setminus D$ would not be  in the class $\Delta_1^1$ because by hypothesis  $L$ is not a $\Delta_1^1$-set. 
But then we could infer from Theorem \ref{eff-ls} that there would exist an $\om$-word $w\in L\setminus D$ having uncountably many accepting runs by 
the B\"uchi Turing machine $(\mathcal{M}, F)$. This is impossible by definition of $D$ and thus we can conclude that $D$ is infinite, i.e. that there 
exist infinitely many  $\om$-words 
which have uncountably many, or equivalently $2^{\aleph_0}$,  accepting runs by $(\mathcal{M}, F)$. 
\qed 

\begin{rem}
We can not obtain a stronger result like ``there  exist   $2^{\aleph_0}$ $\om$-words 
which have $2^{\aleph_0}$ accepting runs by $(\mathcal{M}, F)$" in the conclusion of the above Theorem \ref{inf} 
because there are some countable subsets of $\Sio$ which are in the class 
$\Si_1^1 \setminus \Delta_1^1$. 
\end{rem}

\begin{rem}
The result given by Theorem \ref{inf} is a  dichotomy result for recursive $\om$-languages. A recursive $\om$-language $L$ 
 is either unambiguous or has a great degree of ambiguity: for every B\"uchi Turing machine $(\mathcal{M}, F)$  accepting it there exist infinitely many  
$\om$-words which have $2^{\aleph_0}$ accepting runs by $(\mathcal{M}, F)$. This could  be compared to the case of  context-free $\om$-languages 
accepted by B\"uchi pushdown automata: it is proved in  \cite{Fin03b} that there exist some context-free $\om$-languages which are inherently ambiguous 
of every finite degree $n\geq 2$ (and also some others of infinite degree). 
\end{rem}

There are  many examples of   recursive $\om$-languages which are Borel  and inherently ambiguous of great degree 
since  there are some sets which are 
$(\Si_1^1 \setminus \Delta_1^1)$-sets  in every Borel class 
${\bf \Si}^0_\alpha$ or ${\bf \Pi}^0_\alpha$.  On the other hand recall that 
Kechris, Marker and Sami proved in \cite{KMS89} that the supremum 
of the set of Borel ranks of  (effective) $\Si_1^1$-sets is the ordinal $\gamma_2^1$. 
 This ordinal is precisely defined in \cite{KMS89} where it is  proved to be 
  strictly greater than the ordinal $\delta_2^1$ which is the first non-$\Delta_2^1$ ordinal. 
In particular it holds that $\om_1^{\mathrm{CK}} < \gamma_2^1$, where   $\om_1^{\mathrm{CK}}$ is the first non-recursive ordinal. 
On the other hand it is known that the ordinals  $\gamma < \om_1^{\mathrm{CK}}$ are  the Borel ranks 
of   (effective) $\Delta_1^1$-sets.  Thus we can state the following result. 

\begin{prop}
If  $\Si$ is a finite alphabet and  $L \subseteq  \Sio$  is a recursive $\om$-language which is Borel of  rank $\alpha$  greater than or equal to  the ordinal  
$\om_1^{\mathrm{CK}}$ then for every B\"uchi Turing machine $(\mathcal{M}, F)$  accepting it there exist infinitely many  
$\om$-words which have $2^{\aleph_0}$ accepting runs by $(\mathcal{M}, F)$. 
\end{prop}
 
Notice that this can be applied in a similar way to context-free $\om$-languages accepted by B\"uchi  pushdown automata 
and to infinitary rational relations accepted by B\"uchi 2-tape automata, where ambiguity refers here to acceptance by these less powerful 
accepting devices, see \cite{Fin03b,Fink-Sim}. 
If  $L \subseteq  \Sio$  is a context-free $\om$-language (respectively, $L \subseteq  \Sio \times \Gao$  is an infinitary rational relation) 
which is Borel of  rank $\alpha$  greater than or equal to  the ordinal  
$\om_1^{\mathrm{CK}}$ then $L$ is an inherently ambiguous context-free $\om$-language (respectively,  infinitary rational relation) of degree 
$2^{\aleph_0}$ as defined in  \cite{Fin03b,Fink-Sim}. 

\hs We have established  in Theorem \ref{mainthe} that if $L \subseteq  \Sio$ is an $\om$-language accepted by a 
B\"uchi Turing machine $(\mathcal{M}, F)$   such that $L$ 
is an analytic but non-Borel set, then the set of $\om$-words, 
which have $2^{\aleph_0}$ accepting runs by $(\mathcal{M}, F)$, has cardinality $2^{\aleph_0}$. It is then very natural to ask whether this 
very strong ambiguity property is characteristic of  {\it  non-Borel} recursive $\om$-languages or if some {\it  Borel} recursive $\om$-languages could also 
have this strongest degree of ambiguity.  We first formally define this notion.

\begin{defi}
Let $\Si$ be a finite alphabet and  $L\subseteq \Sio$ be a recursive $\om$-language. 
Then the $\om$-language $L$ is said to have the maximum degree of ambiguity if,  for every B\"uchi Turing machine $(\mathcal{M}, F)$ accepting $L$,  
the set of $\om$-words, 
which have $2^{\aleph_0}$ accepting runs by $(\mathcal{M}, F)$, has cardinality $2^{\aleph_0}$. The set of recursive $\om$-languages having the 
 maximum degree of ambiguity is denoted {\rm Max-Amb}. 
\end{defi}

We are firstly going to state some  undecidability properties. Recall that a B\"uchi Turing machine has a finite description and thus one can associate in 
a recursive and injective manner a positive integer $z$ to each B\"uchi Turing machine $\mathcal{T}$. The integer $z$ is then called the index of the machine 
$\mathcal{T}$.  In the sequel we consider we have fixed such a G\"odel numbering 
of the B\"uchi Turing machines, as in \cite{Fin-ICST,Fin-HI}, and  the  B\"uchi Turing machine  of index $z$, 
reading words over the alphabet $\Ga=\{a, b\}$,  will be denoted $\mathcal{T}_z$. 

 We  recall the notions of 1-reduction and of    $\Sigma^1_n$-completeness (respectively,           $\Pi^1_n$-completeness) for subsets of 
$\mathbb{N}$ (or of $\mathbb{N}^l$ for some integer $l\geq 2$). 
Given two sets $A,B \subseteq \mathbb{N}$ we say $A$ is 1-reducible to $B$ and write $A \leq_1 B$
if there exists a total computable injective  function $f$ from      $\mathbb{N}$     to   $\mathbb{N}$        with $A = f ^{-1}[B]$. 
A set $A \subseteq \mathbb{N}$ is said to be $\Sigma^1_n$-complete   (respectively,   $\Pi^1_n$-complete)  iff $A$ is a  $\Sigma^1_n$-set 
 (respectively,   $\Pi^1_n$-set) and for each $\Sigma^1_n$-set  (respectively,   $\Pi^1_n$-set) $B \subseteq \mathbb{N}$ it holds that 
$B \leq_1 A$. It is known that, for each integer $n\geq 1$, there exist some $\Sigma^1_n$-complete and some $\Pi^1_n$-complete   subsets 
 of $\mathbb{N}$; some examples of such sets are  described in \cite{rog, cc}. 

\begin{thm}\label{unamb}
\noi  The unambiguity  problem      for $\om$-languages of  B\"uchi Turing machines  is 
 $\Pi_2^1$-complete, i.e. :  ~~ The set  $ \{  z \in \mathbb{N}  \mid  L(\mathcal{T}_z) \mbox{ is non-ambiguous  }\}\mbox{  is } \Pi_2^1\mbox{-complete.}$
\end{thm}

\proof We can first express 
``$\mathcal{T}_z$ is non-ambiguous"  by : 
$$``\fa x \in \Ga^\om ~ \fa r, r' \in \{0, 1\}^\om  [(  r \mbox{ and } r' \mbox{ are accepting runs of  } \mathcal{T}_z  \mbox{ on } x) \ra  r=r']"$$
\noi which is a $\Pi_1^1$-formula. 
 Then ``$L(\mathcal{T}_z) $ is non-ambiguous" can be expressed by the following formula:  
 ``$\exists y [ L(\mathcal{T}_z) = L(\mathcal{T}_y)  \mbox{ and }  ( \mathcal{T}_y   \mbox{ is non-ambiguous} ) ]$". This is a $\Pi_2^1$-formula because 
``$L(\mathcal{T}_z) = L(\mathcal{T}_y)$" can be expressed by the  $\Pi_2^1$-formula  
$$``\fa x \in \Ga^\om ~ [ ( x\in L(\mathcal{T}_z) \mbox{ and }  x\in L(\mathcal{T}_y)  ) \mbox{ or  } 
 ( x\notin L(\mathcal{T}_z) \mbox{ and }  x\notin L(\mathcal{T}_y)  ) ]" ,$$ 
\noi and the quantification  $\exists y$ is a first-order quantification bearing on integers. 
Thus the set $\{  z \in \mathbb{N}  \mid  L(\mathcal{T}_z) \mbox{ is non-ambiguous  }\}$  is  a $\Pi_2^1$-set.

\hs To prove completeness we  use a construction we already used in   \cite{Fin-HI}. 
We first define the following operation on $\om$-languages. For $x, x' \in \Gao$ the $\om$-word $x \otimes x'$ is defined by: 
for every integer $n\geq 1$ ~$(x \otimes x')(2n -1)=x(n)$ and $(x \otimes x')(2n)=x'(n)$. 
For two $\om$-languages $L, L' \subseteq \Gao$, the $\om$-language $L \otimes L' $ is defined by $L \otimes L' = \{ x \otimes x' \mid x\in L \mbox{ and } 
x'\in L' \}$. 

We  know that there is a simple example of ${\bf \Si}^1_1$-complete set $L \subseteq \Gao$  accepted by a  B\"uchi Turing machine.  
It  is then easy to define an injective computable function $\theta$ from $\mathbb{N}$ into $\mathbb{N}$ such that, for every integer $z \in \mathbb{N}$, 
it holds that $L(\mathcal{T}_{\theta(z)}) = (L  \otimes   \Gao)  \cup     (\Gao    \otimes   L(\mathcal{T}_{z}))$.  
There are now two cases. 
\nl {\bf First case.}  $L(\mathcal{T}_z)= \Ga^\om$. Then $L(\mathcal{T}_{\theta(z)}) = \Ga^\om$ and $L(\mathcal{T}_{\theta(z)})$ is unambiguous. 
\nl {\bf Second case.} $L(\mathcal{T}_z) \neq  \Ga^\om$. Then there is an $\om$-word $x \in \Gao$ such that $x \notin L(\mathcal{T}_z)$. But 
$L(\mathcal{T}_{\theta(z)}) = (L  \otimes   \Gao)  \cup     (\Gao    \otimes   L(\mathcal{T}_{z}))$ thus 
$\{ \sigma \in \Gao \mid \sigma \otimes x \in L(\mathcal{T}_{\theta(z)}) \} = L$ is a ${\bf \Si}^1_1$-complete set. 
Thus $L(\mathcal{T}_{\theta(z)}) $ is not Borel and 
this implies, by Theorem \ref{mainthe},  that $L(\mathcal{T}_{\theta(z)}) $ is in {\rm Max-Amb} and in particular that 
  $L(\mathcal{T}_{\theta(z)}) $ is inherently 
ambiguous. 

We have  proved,  using the reduction $\theta$, that : 
 $$\{ z \in \mathbb{N} \mid  L(\mathcal{T}_z) = \Ga^\om \} \leq_1 \{ z \in \mathbb{N} \mid  L(\mathcal{T}_z)   \mbox{ is non-ambiguous  } \}$$ 
\noi Thus this latter set is  $\Pi_2^1$-complete because the universality problem for $\om$-languages of Turing machines is itself $\Pi_2^1$-complete, 
see \cite{cc, Fin-HI}. 
\qed 

\hs 

\begin{thm}\label{maxamb}
The set  $\{  z \in \mathbb{N}  \mid  L(\mathcal{T}_z) \in {\rm Max}\mbox{-}{\rm Amb}  \}\mbox{  is } \Si_2^1\mbox{-complete.}$
\end{thm}

\proof We first show that the set $\{  z \in \mathbb{N}  \mid  L(\mathcal{T}_z) \in {\rm Max}\mbox{-}{\rm Amb} \}$ is in the class $\Si_2^1$. 
In a similar way as  in the proof of Proposition \ref{p1} we consider the set 
$R_z$  defined  by: 
$$R_z= \{ (w,  r) \mid w\in \Ga^\om \mbox{ and }  r \in \{0, 1\}^\om  \mbox{ is an accepting run of } \mathcal{T}_z  \mbox{ on  the $\om$-word } w \}.$$
\noi This set $R_z$ is a $\Delta^1_1$-subset of  $\Ga^\om \times   \{0, 1\}^\om$. 
Notice that  the set of accepting runs of $\mathcal{T}_z$ on  an $\om$-word $w\in \Ga^\om$ is the section 
$$R_{z, w}=\{  r \in \{0, 1\}^\om \mid (w,  r) \in R_z\}.$$
\noi It is a set in the relativized class  $\Si_1^1(w)$ and thus it is        uncountable iff it contains a point $r_0$ such that $\{r_0\}$ is not a 
$\Delta^1_1(w)$-subset of  $\{0, 1\}^\om$. Moreover we have already seen that the set 
$$D_z=\{  w \mid w\in \Ga^\om \mbox{ and there are  uncountably many  accepting runs of } \mathcal{T}_z  \mbox{ on } w \}.$$
\noi is a  $\Si_1^1$-set. Thus it is uncountable iff it contains a member which is not in class $\Delta^1_1$. 
Recall now that Harrington, Kechris and Louveau obtained  a coding of $\Delta_1^1$-subsets  (respectively, of  $\Delta_1^1(w)$-subsets) 
of  $\{0, 1\}^\om$  in \cite{HKL}, (see the proof of the 
above Theorem \ref{inf}). Then there is a $\Pi_1^1$-formula $\Theta_1(w)$ such that for every $w\in \Ga^\om$ it holds that $\{w\}$ is in the class $\Delta_1^1$ iff  
$\Theta_1(w)$ holds. And there is a $\Pi_1^1$-formula $\Theta_2(w, r)$ such that for every $w\in \Ga^\om$ and $ r \in \{0, 1\}^\om$ 
it holds that $\{r\}$ is in the class $\Delta_1^1(w)$ iff  $\Theta_2(w, r)$ holds. We can now express the sentence ``the set of  $\om$-words, 
which have $2^{\aleph_0}$ accepting runs by $\mathcal{T}_z$, has cardinality $2^{\aleph_0}$" by the following formula  $\Omega(z)$: 

$$\exists w \exists r [ \neg \Theta_1(w)  \wedge  \neg \Theta_2(w, r)   \wedge  (w,  r) \in R_z  ]$$
\noi This formula $\Omega(z)$ is clearly a $\Sigma_1^1$-formula. We can now express the sentence  $``L(\mathcal{T}_z) \in {\rm Max}\mbox{-}{\rm Amb}"$
by the following sentence: 
$$\fa z' \in \mathbb{N} [ L(\mathcal{T}_z) \neq  L(\mathcal{T}_{z'})  \mbox{ or }  \Omega(z') ] $$ 
\noi This is a $\Si_2^1$-formula because ``$L(\mathcal{T}_z) \neq  L(\mathcal{T}_{z'})$" is easily espressed by a $\Si_2^1$-formula 
(see the proof of Theorem \ref{unamb}), the formula $\Omega(z)$ is a $\Sigma_1^1$-formula, and the first-order quantification $\fa z'$ bears on integers. 
Thus we have proved that the set $\{  z \in \mathbb{N}  \mid  L(\mathcal{T}_z) \in {\rm Max}\mbox{-}{\rm Amb} \}$ is in the class $\Si_2^1$. 

\hs To prove the completeness part of the theorem we can use the same reduction $\theta$ as in the proof of the preceding theorem. 
Recall that we  know that there is a simple example of ${\bf \Si}^1_1$-complete set $L \subseteq \Gao$  accepted by a  B\"uchi Turing machine.  
We have  defined, in the proof of the preceding theorem,  an injective computable function $\theta$ from $\mathbb{N}$ into $\mathbb{N}$ such that,
 for every integer $z \in \mathbb{N}$, 
it holds that $L(\mathcal{T}_{\theta(z)}) = (L  \otimes   \Gao)  \cup     (\Gao    \otimes   L(\mathcal{T}_{z}))$.  
We have seen that there are  two cases. 

\hs \noi {\bf First case.}  $L(\mathcal{T}_z)= \Ga^\om$. Then $L(\mathcal{T}_{\theta(z)}) = \Ga^\om$ and $L(\mathcal{T}_{\theta(z)})$ is unambiguous. 
\nl {\bf Second case.} $L(\mathcal{T}_z) \neq  \Ga^\om$. Then $L(\mathcal{T}_{\theta(z)}) $ is in {\rm Max-Amb}. 

\hs Thus we have  proved,  using the reduction $\theta$, that : 
 $$\{ z \in \mathbb{N} \mid  L(\mathcal{T}_z) \neq \Ga^\om \} \leq_1 \{ z \in \mathbb{N} \mid  L(\mathcal{T}_z)   \mbox{ is in {\rm Max-Amb}  } \}$$ 
\noi Thus this latter set is  $\Si_2^1$-complete because the universality problem for $\om$-languages of Turing machines is itself $\Pi_2^1$-complete, 
see \cite{cc, Fin-HI}, so $\{ z \in \mathbb{N} \mid  L(\mathcal{T}_z) \neq \Ga^\om \}$ is $\Si_2^1$-complete. 
\qed 

\hs We now briefly recall some notions of set theory which will be useful for the next result and
refer the reader to a textbook like \cite{Jech} for more background on set theory. 

  The usual axiomatic system ZFC is 
Zermelo-Fraenkel system ZF   plus the axiom of choice  AC. 
 The axioms of ZFC express some  natural facts that we consider to hold in the universe of sets. 
A model ({\bf V}, $\in)$ of  an arbitrary set of axioms $\mathbb{A}$  is a collection  {\bf V} of sets,  equipped with 
the membership relation $\in$, where ``$x \in y$" means that the set $x$ is an element of the set $y$, which satisfies the axioms of   $\mathbb{A}$. 
We  often say `` the model {\bf V}" instead of "the model ({\bf V}, $\in)$". 

We say that two sets $A$ and $B$ have same cardinality iff there is a bijection from $A$ onto $B$ and we denote this  by $A \approx B$. 
The relation $\approx$ is an equivalence relation. 
Using the axiom of choice AC, one can prove that any set $A$ can be well-ordered so  there is an ordinal $\gamma$ such that $A \approx \gamma$. 
In set theory the cardinality of the set $A$ is then formally defined as the smallest such ordinal $\gamma$. Such ordinals $\gamma$ are also called cardinal 
numbers, or simply cardinals. 
 The infinite cardinals are usually denoted by
$\aleph_0, \aleph_1, \aleph_2, \ldots , \aleph_\alpha, \ldots$
The continuum hypothesis  CH  says that the first uncountable cardinal $\aleph_1$ is equal to $2^{\aleph_0}$ which is the cardinal of the 
continuum. 

  If  {\bf V} is  a model of ZF and ${\bf L}$ is  the class of  {\it constructible sets} of   {\bf V}, then the class  ${\bf L}$    is a model of  
 ZFC + CH.
Notice that the axiom  V=L, which  means ``every set is constructible",   is consistent with  ZFC  because   ${\bf L}$ is a model of 
ZFC + V=L, see \cite[pages 175-200]{Jech}. 

 Consider now a model {\bf V} of  ZFC  and the class of its constructible sets ${\bf L} \subseteq {\bf V}$ which is another 
model of  ZFC.  It is known that 
the ordinals of {\bf L} are also the ordinals of  {\bf V}, but the cardinals  in  {\bf V}  may be different from the cardinals in {\bf L}. 
 In particular,  the first uncountable cardinal in {\bf L}  is denoted 
 $\aleph_1^{\bf L}$, and it is in fact an ordinal of {\bf V} which is denoted $\om_1^{\bf L}$. 
  It is well-known that  this ordinal satisfies the inequality 
$\om_1^{\bf L} \leq \om_1$.  In a model {\bf V} of  the axiomatic system  ZFC + V=L the equality $\om_1^{\bf L} = \om_1$ holds, but in 
some other models of  ZFC the inequality may be strict and then $\om_1^{\bf L} < \om_1$. 

The following result was proved in \cite{Fin-ICST}. 

\begin{thm}\label{the4}
 There exists a real-time $1$-counter B\"uchi automaton $\mathcal{A}$, which can be effectively  constructed, such that the topological complexity of the 
$\om$-language $L(\mathcal{A})$ is not determined by the axiomatic system {\rm ZFC}. Indeed it holds that : 
\begin{enumerate}
\item[(1)] {\rm(ZFC $+$ V=L)}.\ The $\om$-language $L(\mathcal{A})$ is an analytic but non-Borel  set. 
\item[(2)] {\rm(ZFC $+$ $\om_1^{\bf L} < \om_1$)}.\ The $\om$-language $L(\mathcal{A})$ is a  ${\bf \Pi}^0_2$-set. 
\end{enumerate}
\end{thm}

We can now show that  it is consistent with ZFC that some recursive $\om$-languages in the Borel class ${\bf \Pi}^0_2$, 
hence  of a low Borel rank,  have the maximum degree of ambiguity.

\begin{thm}\label{the5}
{\rm(ZFC $+$ $\om_1^{\bf L} < \om_1$)}.   There exists an $\om$-language accepted by a real-time $1$-counter B\"uchi automaton 
which belongs to the Borel class ${\bf \Pi}^0_2$ and which has the maximum degree of ambiguity with regard to acceptance by Turing machines, i.e. 
which belongs to the class {\rm Max-Amb}. 
\end{thm}

\proof Consider the real-time $1$-counter B\"uchi automaton $\mathcal{A}$ given by Theorem \ref{the4}. It may be seen as a Turing machine which has an index $z_0$ so that 
$L(\mathcal{A})=L(\mathcal{T}_{z_0})$. 
Let now ${\bf V}$ be a model of ({\rm ZFC} + $\om_1^{\bf L} < \om_1$). In this model $L(\mathcal{A})$  is a Borel set in the class   ${\bf \Pi}^0_2$. 
We are going to show that it is also in the class {\rm Max-Amb}. 

 Consider  the model ${\bf L}$ which is  the class of  {\it constructible sets} of   {\bf V}. The class ${\bf L}$ is a  model  of ({\rm ZFC + V=L}) and thus by 
Theorem \ref{the4} the $\om$-language $L(\mathcal{A})$ is an analytic but non-Borel  set in ${\bf L}$. Then it follows from Theorem \ref{mainthe} that in  ${\bf L}$  
the $\om$-language $L(\mathcal{T}_{z_0})$  is in the class {\rm Max-Amb}. 
On the other hand, the set  $\{ z \in \mathbb{N}  \mid  L(\mathcal{T}_z) \in {\rm Max}\mbox{-}{\rm Amb}  \}$ is a $\Si_2^1$-set by Theorem \ref{maxamb}. Thus by 
the Shoenfield's Absoluteness Theorem (see \cite[page 490]{Jech}) this set is the same in the model ${\bf V}$ {\it and } in the model ${\bf L}$. 
This implies that the  $\om$-language 
$L(\mathcal{A})=L(\mathcal{T}_{z_0})$ has the maximum degree of ambiguity with regard to acceptance by Turing machines in the model ${\bf V}$ too. 
\qed

\begin{rem}
In order to prove Theorem \ref{the5} we do not need to use any large cardinal axiom  or even the consistency of such an axiom, 
because it is known that ({\rm ZFC} + $\om_1^{\bf L} < \om_1$) is equiconsistent with {\rm ZFC}. However it is known that the existence of a measurable cardinal (or even of a larger 
one),  or the axiom of analytic determinacy,  imply the strict inequality $\om_1^{\bf L} < \om_1$ and thus the existence of the 
$\om$-language in the class  {\rm Max-Amb}  given by  Theorem \ref{the5}. 
\end{rem}

\section{Concluding remarks}

We have investigated the notion of ambiguity for recursive $\om$-languages. In particular Theorem \ref{inf} 
gives a remarkable dichotomy result for recursive $\om$-languages: a recursive $\om$-language $L$ 
 is either unambiguous or has a great degree of ambiguity. 

On the other hand,   Theorem \ref{the5} states that it is consistent with ZFC that there exists a recursive  $\om$-language 
which belongs to the Borel class ${\bf \Pi}^0_2$ and which has the maximum degree of ambiguity. The following  question  now naturally arises: 
``Does there exist such a 
recursive  $\om$-language  in {\it every model of } ZFC ?" 

\hs \noi {\bf Acknowledgements.}  We  thank the anonymous referees for their very useful remarks on a preliminary version of this  paper.

\vspace{-40 pt}
\end{document}